\documentstyle [12pt,epsf,worldsci]{article}

\global\arraycolsep=2pt%
\def\coeff	#1#2{{\textstyle{#1\over #2}}}%
\def\half	{\coeff {1}{2}}%
\def\lsim	{\,\,\vcenter{\hbox{$\buildrel{\displaystyle <}\over\sim$}}\,\,}
\def\gsim	{\,\,\vcenter{\hbox{$\buildrel{\displaystyle >}\over\sim$}}\,\,}
\def\sym	{_{\rm sym}}
\def\asym	{_{\rm asym}}
\def\eps	{\epsilon}
\def\mh		{m_{\rm h}}

\begin {document}
\begin {flushright}
UW-PT-94-14
\\October 1994
\end {flushright}
\begin {title}
    {%
    The Electroweak Phase Transition, Part II:
    {\bf $\epsilon$-Expansion Results}
    \footnote
	{
	Based on a talk presented at the Quarks-94 conference in
	Vladimir, Russia, May 1994.
	This work was performed in collaboration with
	P. Arnold and is described in greater
	detail in reference \citenum {A&Y}.
	}%
    }%
\end {title}

\author
    {%
    Laurence G. Yaffe%
    \thanks
	{
	Research supported in part by DOE
	grant DE-FG06-91ER40614.
	}\\
    {\em Department of Physics, FM-15, University of Washington\\
    Seattle, Washington 98195 USA}
    }
\maketitle

\bigskip
\baselineskip 14pt

    Detailed knowledge of the electroweak phase transition is
needed to determine the viability of electroweak baryogenesis.
However, as discussed in the preceding talk by Peter Arnold,
standard perturbative (or mean field theory) techniques are
only adequate for studying the finite-temperature electroweak phase transition
when the Higgs mass is sufficiently small.

    The $\epsilon$-expansion, based on dimensional continuation from
3 to $4{-}\epsilon$ spatial dimensions, provides an alternative
systematic approach for computing the effects of (near)-critical
fluctuations.
After reviewing the basic strategy of the $\epsilon$-expansion
and its application in simple scalar theories,
I will discuss the application of the $\epsilon$-expansion
to electroweak theory,
describe the computation of a variety of physical quantities,
and summarize the results of several tests of the validity of
$\epsilon$-expansion calculations in electroweak theory.%
\footnote
    {%
    See reference \citenum {Others} for other discussions of the
    $\epsilon$-expansion in the context of electroweak theory.
    }

\section {The $\epsilon$-expansion in scalar theory}

    The $\epsilon$-expansion is based on the idea that
instead of trying to solve a theory directly in three spatial dimensions,
it can be useful to generalize the theory
from three to $4{-}\epsilon$ spatial dimensions,
solve the theory near four dimensions (when $\epsilon \ll 1$),
and then extrapolate to the physical case of 3 spatial dimensions.
Specifically, one expands physical quantities in powers of $\epsilon$
and then evaluates the resulting (truncated) series at $\epsilon = 1$
\cite {Wilson}.
This can provide a useful approximation when the
relevant long distance fluctuations are weakly coupled near 4 dimensions,
but become sufficiently strongly coupled that the loop expansion parameter
is no longer small in three dimensions.

    Scalar $\phi^4$ theory (or the Ising model) is a classic example.
In four dimensions, the long distance structure of a quartic scalar
field theory is trivial;
this is reflected in the fact that the renormalization group equation
$
    \mu (d\lambda / d\mu) = \beta_0 \, \lambda^2 + O(\lambda^3) \,,
$
has a single fixed point at $\lambda = 0$.
In $4{-}\epsilon$ dimensions, the canonical dimension of the field changes
and the renormalization group equation acquires a linear term,
$$
    \mu {d\lambda \over d\mu} = -\epsilon \lambda + \beta_0 \, \lambda^2 +
    O(\lambda^3) \,.
$$
This has a non-trivial fixed point
(to which the theory flows as $\mu$ decreases)
at $\lambda^* = \epsilon / \beta_0 + O(\epsilon^2)$.
The fixed point coupling is $O(\epsilon)$ and thus small
near four dimensions, but grows with decreasing dimension and
becomes order one when $\epsilon = 1$.
Near four dimensions,
a perturbative calculation in powers of $\lambda$ is reliable
and directly generates an expansion in powers of $\epsilon$.

The existence of an infrared-stable fixed point indicates the
presence of a continuous phase transition as the
bare parameters of the theory are varied.
Interesting physical quantities include the critical exponents
which characterize the non-analytic behavior at the transition.
Performing conventional (dimensionally regularized) perturbative
calculations of the appropriate anomalous dimensions,
and evaluating the perturbative series at the fixed point,
one finds, for example, that the susceptability exponent
(equivalent to the anomalous dimension of $\phi^2$) has the expansion
\cite {Wilson,Gorishny}
\begin {equation}
\label {Igamma}
   \gamma = 1 + 0.167 \, \epsilon + 0.077 \, \epsilon^2 - 0.049 \, \epsilon^3
   + O(\epsilon^4) \,,
\end {equation}
while the exponent characterizing the power-law decay of the
propagator at the critical point
(equivalent to the anomalous dimension of $\phi$) is
\cite {Wilson,Chetyrkin}
\begin {equation}
\label {Ieta}
    \eta = 0.0185 \, \epsilon^2 + 0.0187 \, \epsilon^3
    - 0.0083 \, \epsilon^4 + 0.0359 \, \epsilon^5 + O(\epsilon^6) \,.
\end {equation}
Adding the first three non-trivial terms in these series,
and evaluating at $\epsilon=1$,
yields results which agree quite well with the best available results
for these exponents.
(For $\gamma$, the $\epsilon$-expansion gives $1.195$, to be compared with
$1.2405 \pm 0.0015$, while for $\eta$ one finds
0.029 versus 0.035 \cite {Zinn-Justin,Gupta}.)

Inevitably, perturbative expansions in powers of $\lambda$
are only asymptotic;
coefficients grow like $n!$, so that succeeding terms in the series
begin growing in magnitude when
$
    n \gsim O(1/\lambda)
$.
Expansions in $\epsilon$ are therefore also asymptotic,
with terms growing in magnitude beyond some order
$n \gsim O(1/\epsilon)$.
If one is lucky, as is the case in the pure scalar theory,
$O(1/\epsilon)$ really means something like three or four
when $\epsilon = 1$ and the first few terms of the series will be useful.
If one is unlucky, no terms in the expansion will be useful.
Whether or not one will be lucky cannot be determined in advance of an
actual calculation.

\section {Electroweak Theory}

    To apply the $\epsilon$-expansion to electroweak theory,
one begins with the full $3{+}1$ dimensional finite temperature
Euclidean quantum field theory (in which one dimension is periodic
with period $\beta = 1/T$) and integrates out all non-static Fourier
components of the fields.
The integration over modes with momenta of order $T$ or larger
may be reliably performed using standard perturbation theory in
the weakly-coupled electroweak theory.
This reduces the theory to an effective 3-dimensional SU(2)-Higgs theory
with a renormalization point $\mu$ which may be conveniently chosen to
equal the temperature.%
\footnote
    {%
    Fermions, having no static Fourier components, are completely
    eliminated in the effective theory.
    For simplicity, the effects of a non-zero weak mixing angle
    and the resulting perturbations due to the U(1) gauge field
    are ignored.
    Finally, one may also integrate out the static part of
    the time component of the gauge field, since this field
    acquires an $O(gT)$ Debye-screening mass.
    }
The effective theory depends on three relevant renormalized parameters:
\begin {center}
\begin {tabular}{rl}
    $g_1(T)^2_{\strut}$		\quad---& the SU(2) gauge coupling,\\
    $\lambda_1(T)_{\strut}$	\quad---& the quartic Higgs coupling,\\
    $m_1(T)^2$			\quad---& the Higgs mass (squared).
\end {tabular}
\end {center}

\begin {figure}
\vbox
    {%
    \begin {center}
	\leavevmode
	
	\epsfbox [72 260 520 550] {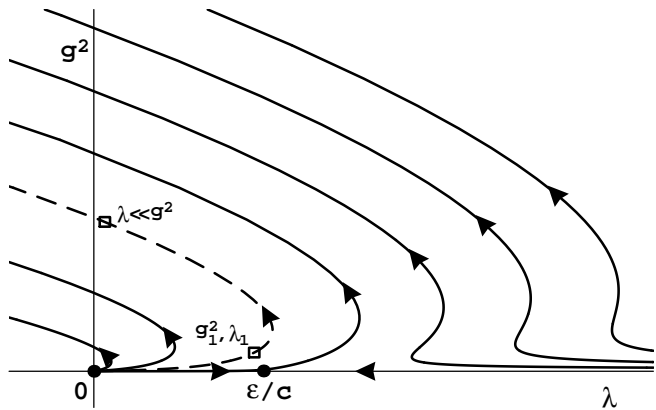}
    \end {center}
    \caption
	{%
	\label {figc}
	The renormalization group flow for an SU(2)-Higgs theory.
	Arrows indicate the direction of decreasing renormalization point.
	The dashed line is the trajectory which flows from an initial set
	of couplings $(g_1^2, \lambda_1)$ into the region where
	$\lambda \ll g^2$.
	}%
    }%
\end {figure}

    Next, one replaces the 3-dimensional theory by the corresponding
$4{-}\epsilon$ dimensional theory (and scales the couplings so that
$g_1^2 / \epsilon$ and $\lambda_1 / \epsilon$ are held fixed).
This is the starting point for the $\epsilon$-expansion.
When $\epsilon$ is small, one may reliably compute the
renormalization group flow of the effective couplings.
The renormalization group equations have the form
\begin {eqnarray}
    \mu {d\lambda \over d\mu}
&=&
    -\epsilon \, \lambda
    + (a \, g^4 + b \, g^2 \lambda + c \, \lambda^2) + \cdots \,,
\label {RG-a}
\\
    \mu {dg^2 \over d\mu}
&=&
    -\epsilon \, g^2
    + \beta_0 \, g^4 + \cdots \,.
\label {RG-b}
\end {eqnarray}
The precise values of the coefficients (and the next order terms)
may be found in reference \citenum {A&Y}.
These equations may be integrated analytically, and produce the
flow illustrated in figure \ref {figc}.

    Note that a non-zero gauge coupling renders the Ising fixed point
at $\lambda = \beta_0 / \epsilon$ unstable, and that no other
(weakly coupled) stable renormalization group fixed point exists.
Trajectories with $g^2 > 0$ eventually cross the $\lambda = 0$ axis
and flow into the region where the theory (classically) would appear
to be unstable.
Such behavior is typically indicative of a first-order phase transition
\cite {Ginsparg}.
To determine whether this is really the case,
one must be able to perform a reliable calculation of the effective
potential (or other physical observables).
As discussed in Peter Arnold's talk, the loop expansion parameter
for long distance physics is $\lambda(\mu) / g^2(\mu)$.
Consequently, the best strategy is to use the renormalization group
to flow from the original theory at $\mu = T$, which may have
$\lambda(T) / g^2(T)$ large, to an equivalent theory with $\mu \ll T$
for which $\lambda (\mu) / g^2(\mu)$ is small.
This is equivalent to the condition that one decrease the renormalization
point until it is comparable to the relevant scale for long distance physics,
specifically, the gauge boson mass, $M$.
By doing so, one eliminates large factors of
$\left[ (M / \mu)^\epsilon - 1 \right] / \epsilon$
which would otherwise spoil the reliability of the loop expansion.
(This, of course, is nothing other than the transcription to $4{-}\epsilon$
dimensions of the usual story in 4 dimensions, where
appropriate use of the renormalization group allows one to sum up
large logarithms which would otherwise spoil the perturbation expansion.)

    For small $\epsilon$, the change of scale required to flow from
an initial theory where $\lambda_1 / g_1^2 = O(1)$ to an equivalent theory
with $\lambda(\mu) / g^2(\mu) \ll 1$ is exponentially large;
the ratio of scales is
$$
    s \equiv {T \over \mu}
    \sim e^{\lambda_1 / g_1^4}
    \sim e^{O(1/\epsilon)} \,.
$$
This is easy to see directly from the renormalization group equations
(\ref {RG-a}) and (\ref {RG-b}).
Since $g_1^2$ and $\lambda_1$ are (by construction) both $O(\epsilon)$,
all terms on the right-hand side of the renormalization group equations
are $O(\epsilon^2)$.
Hence, a change in $\ln \mu$ of $O(1/\epsilon)$ is required to produce
an order one change in the ratio of $\lambda / g^2$.

    Given the parameters $g^2(\mu)$, $\lambda(\mu)$ and $m^2(\mu)$
of the resulting effective theory, one may use the usual loop expansion
to compute interesting physical quantities.
Because the change in scale is exponentially sensitive to $1/\epsilon$,
the result for a typical physical quantity will have the schematic form
\begin {eqnarray}
    {\cal O}
&=&
    f[g^2(\mu), \lambda(\mu), m^2(\mu)]
    \left( {\mu \over T} \right)^\#
\\
&\sim&
    \epsilon^\# \left( 1 + O(\epsilon) + \cdots \right) \;
    \exp \! \left[
	{\# \over \epsilon} \left( 1 + O(\epsilon) + \cdots \right)
    \right] \,.
\label {expansion}
\end {eqnarray}
In general, a calculation accurate to $O(\epsilon^n)$ requires
an $n$-loop calculation in the final effective theory, together
with $n{+}1$ loop renormalization group evolution.

To obtain predictions for the original theory in three spatial
dimensions, one finally truncates the expansions at a given order
and then extrapolates from $\epsilon \ll 1$ to $\epsilon = 1$.
Just as for the simple $\phi^4$ theory, the reliability of the resulting
predictions at $\epsilon = 1$ can only be tested {\em a-posteriori}.

Peter Arnold and I have carried out the above precedure for
a variety of observables characterizing the electroweak
phase transition at both leading and next-to-leading order in the
$\epsilon$-expansion.
To obtain leading order results, one must first integrate the one-loop
renormalization group equations, determine the change in scale $\mu / T$,
and express the renormalized parameters $g^2(\mu)$ and $\lambda(\mu)$
in terms of the initial parameters $g_1^2$ and $\lambda_1$.
It is convenient to choose the final renormalization point
as precisely that scale where the (minimally subtracted) value
of $\lambda(\mu)$ vanishes.
This greatly simplifies the resulting formula for the effective potential.
Computing the one-loop effective potential of the theory with
$\lambda(\mu) = 0$ is easy; one finds the characteristic
Coleman-Weinberg form
\begin {equation}
    V_{\rm eff}(\phi) \, \mu^\epsilon
    =
    \half m^2 \bar\phi^2 +
    \coeff 14 \, a \, (g\bar \phi)^4
    \left[ \ln \left( {g \bar \phi \over \mu } \right) + {\rm const.} \right].
\end {equation}
As $m^2(\mu)$ (which is a function of $T$) varies, the minimum of the
effective potential jumps discontinuously from the symmetric minimum at
$\phi = 0$ to the asymmetric minimum where $g\phi = O(\mu)$.
In other words, the theory (for $\epsilon \ll 1$) undergoes a first order
phase transition, for all initial values of $\lambda_1 / g_1^2$.
Computing, for example, the scalar correlation length
at the transition in the asymmetric phase yields
\begin {eqnarray}
    \xi\asym
&=&
    {1 \over \mu} \> {\# \over g(\mu) }
    = {1 \over T} \> {\# \over \sqrt \epsilon} \> e^{-\# / \epsilon}
\\
\noalign {\hbox {and}}
    g^2(\mu)
&=&
    s^\epsilon g_1^2 \Bigm/
    \left[
	1 + \beta_0 (s^\epsilon {-} 1) g_1^2 / \epsilon
    \right].
\end {eqnarray}
Explicit values for the unspecified constants above, plus the
explicit (but rather involved) expression for $s^\epsilon$ as a function
of $g_1^2 / \epsilon$ and $\lambda_1 / \epsilon$, may be found in
ref.~\citenum {A&Y}.
Similar lowest-order results were also found for the scalar correlation in
the symmetric phase,
the free energy difference between the symmetric and asymmetric phases
$\Delta F(T)$,
the latent heat $\Delta Q = -T d \Delta F / dT |_{T_c}$,
the surface tension $\sigma$ between symmetric and asymmetric phases at $T_c$,
the bubble nucleation rate $\Gamma_N(T)$ below $T_c$,
and the baryon violation (or sphaleron) rate $\Gamma_B(T_c)$.

    The lowest order $\epsilon$-expansion predictions differ from
the results of standard one-loop perturbation theory (performed directly
in three space dimensions) in several interesting ways.
First, the $\epsilon$-expansion predicts a {\em stronger} first order
transition than does one-loop perturbation theory
(as long as $M_{\rm H} < 130$ GeV).
The correlation length at the transition is smaller, and the latent
heat larger, than the perturbation theory results.
The size of the difference depends on the value of the
Higgs mass; see reference \citenum {A&Y} for quantitative results.
Naively, one would expect that a stronger first order transition
would imply a smaller baryon violation rate (since a larger
effective potential barrier between the co-existing phases should
decrease the likelyhood of thermally-activated transitions
across the barrier).
This expectation is wrong (in essence, because it unjustifiably
assumes that the shape of the barrier remains unchanged).
Along with predicting a strengthing of the transtion,
the $\epsilon$-expansion predicts a {\em larger}
baryon violation rate.
This occurs because the baryon violation rate is exponentially
sensitive to the sphaleron mass ($=$ the electroweak barrier height),
$$
    \Gamma_{\rm B} \propto \exp -S_{\rm sphaleron} \,,
$$
and the sphaleron mass depends inversely on $\epsilon$,
$$
    S_{\rm sphaleron} = {\# \over g^2(\mu)} = O(1/\epsilon) \,.
$$
Hence, unlike other observables, the exponential sensitivity to
$1/\epsilon$ in the baryon violation rate does not arise solely
from an overall power of the scale factor $\mu / T$.

Note that an increase in the baryon violation rate (compared to
standard perturbation theory) will make the constraints
for viable electroweak baryogenesis more stringent;
specifically, the (lowest order) $\epsilon$-expansion suggests that
the minimal standard model bound $M_{\rm H} \lsim 35$--40
GeV derived using one-loop perturbation
theory in ref.~\citenum {Dine} should be even lower,
further ruling out electroweak baryogenesis in the minimal model.

\section {Testing the $\epsilon$-expansion}

    As mentioned earlier, in general there is no way to know,
in advance of an actual calcualtion, how many terms (if any)
in an $\epsilon$-expansion will be useful when results are
extrapolated to $\epsilon = 1$.
Therefore, in order to assess the reliability of $\epsilon$-expansion,
one must try to test predictions for actual physical quantities.
For the electroweak theory, three types of tests are possible.
\begin {itemize}
\item [\bf A.]
    $\lambda \ll g^2$.
    In the limit of a light (zero temperature) Higgs mass,
    or equivalently small $\lambda_1 / g_1^2$,
    the loop expansion in three dimensions is reliable.
    Hence, although this is not a realistic domain, one may easily
    test the reliability of the $\epsilon$-expansion in this regime by
    comparing with direct three-dimensional perturbative calcualtions.
    Table~1 summarizes the fractional error for various
    physical quantities produced by truncating the $\epsilon$-expansion
    at leading, or next-to-leading, order before evaluating at
    $\epsilon = 1$, in the light Higgs limit.
    Although the lowest-order results often error by a factor of two
    or more, all but one of the next-to-leading order results
    are correct to better than 10\%.
    (The free energy difference at the limit of metastability,
    $\Delta F(T_0)$, has the most poorly behaved $\epsilon$-expansion.
    However, if one instead computes the logarithm of this quantity,
    then the next-to-leading order result is correct to within 17\%.
    The baryon violation rate is not shown because, due to the way
    its $\epsilon$-expansion was constructed, the result is trivially
    the same as the three-dimensional answer when $\lambda_1 \ll g_1^2$.
    See ref.~\citenum {A&Y} for details.)
    \begin{table}[b]
    \begin {center}
    \tabcolsep=8pt
    \begin {tabular}{|lc|cc|}             \hline
    \multicolumn{1}{|c}{observable ratio}
      &                   & LO          & NLO         \\ \hline
    asymmetric correlation length
      &   $\xi\asym$      & \phantom-0.14        & -0.06        \\
    symmetric correlation length
      &   $\xi\sym$       & \phantom-0.62        & -0.08        \\
    latent heat
      &   $\Delta Q$      & -0.23        	&\phantom-0.04	\\
    surface tension
      &   $\sigma$        & -0.40        	& -0.02        \\
    free energy difference
      &   $\Delta F(T_0)$ & -0.76        	& -0.44        \\ \hline
    \end {tabular}
    \end {center}
    \caption
	{%
	\label {tableb}
	The fractional error in the $\eps$-expansion results,
	when computing prefactors through leading order (LO) and
	next-to-leading order (NLO) in $\eps$,
	when $\lambda_1 \ll q_1^2$.
	}%
    \end{table}
\item [\bf B.]
    $\lambda \gsim g^2$.
    When $\lambda / g^2$ is $O(1)$, the three-dimensional loop expansion
    is no longer trustworthy.
    However, one may still test the stability of $\epsilon$-expansion
    predictions by comparing $O(\epsilon^n)$ and $O(\epsilon^{n+1})$
    predictions --- provided, of course, one can evaluate at least two
    non-trivial orders in the $\epsilon$-expansion.
    For most physical quantities this is not (yet) possible;
    determining the lowest-order behavior of the prefactor in
    expansion (\ref {expansion}) requires a one-loop calculation
    in the final effective theory together with a two-loop evaluation
    of the (solution to the) renormalization group equations.
    A consistent next-to-leading order calculation requires a two-loop
    calculation in the final theory together with three-loop
    renormalization group evolution.
    Althouth two-loop results for the effective potential and
    beta functions are known, three loop renormalization group
    coefficients in the scalar sector are not currently available.
    Nevertheless, by taking suitable combinations of physical quantities
    one can cancel the leading dependence on the scale ratio
    $\mu / T$ and thereby eliminate the dependence
    (at next-to-leading order) on the three loop beta functions.
    For example, the latent heat depends on the scale as
    $\Delta Q \sim (\mu / T)^{2+\epsilon}$ while the
    scalar correlation length $\xi \sim (\mu / T)^{-1}$.
    Therefore, the combination $\xi^2 \Delta Q$ cancels the leading
    $\mu / T \sim e^{O(1/\epsilon)}$ dependence and thus requires
    only two-loop information for its next-to-leading order evaluation.
    The result of the (rather tedious) calculation may be put in
    the form
    \begin {equation}
	\xi\asym^2 \Delta Q
	=
	T^{1-\epsilon} \, f(f_1^2, \lambda_1)
	\left[ 1 + \delta + O(\epsilon^2) \right] \,,
    \end {equation}
    where $\delta$, the relative size of the next-to-leading order
    correction, is plotted in figure 2.
    The correction varies between roughly $\pm30\%$ for
    (zero temperature) Higgs masses up to 150 GeV.
    This suggests that the $\epsilon$ expansion is tolerably
    well behaved for these masses.
    For larger masses the correction does not grow indefinitely,
    but is bounded by 80\%,
    suggesting that the $\epsilon$ expansion may remain
    qualitatively useful even when it does not work as well
    quantitatively.
    \begin {figure}[h]
    \vbox
	{%
	\begin {center}
	    \leavevmode
	    
	    \epsfbox [150 250 500 500] {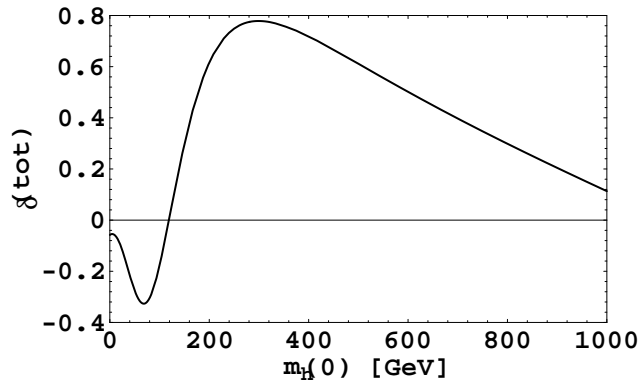}
	\end {center}
	\caption
	    {%
	    \label {figlat}
	    The relative size of the next-to-leading order correction
	    to $\xi\asym^2 \Delta Q$ in the $\eps$-expansion.  The values
	    are given as a function of the (tree-level)
	    zero-temperature Higgs mass in minimal SU(2)
	    theory ($N=2$) with $g = 0.63$.
	    }%
	}%
    \vspace* {-.1in}
    \end {figure}

    The comparatively small size of $\delta$
    becomes more impressive when one
    examines the size of the different pieces which contribute.
    One may separate the effects produced by the second order
    shift in the mass parameter of the effective theory
    $\delta m^2(\mu)$,
    the shift in the effective gauge coupling $\delta g^2(\mu)$,
    the change in scale $\delta (\mu/T)$,
    and the change in the final evaluation of the latent heat.
    At $\mh(0) = 80$ and 250 GeV, the four different contributions are
    \begin {eqnarray}
       \delta\,(\phantom{2} 80~\hbox{GeV})
	  &=& -0.45 + \phantom{3} 4.88 - \phantom{2} 0.94 - 3.79
	  = -0.30 \,,
    \\
       \delta\,(250~\hbox{GeV})
	  &=& -4.42 + 34.83 - 20.00 - 9.66
	  = \phantom{-} 0.75 \,,
    \end {eqnarray}
    respectively.
    These large cancellations clearly underscore the importance of examining
    {\it physical\/} quantities, rather than unphysical ones such as
    $q^2(s)$ or $s^\eps$, when testing the $\eps$-expansion.
\item [\bf C.]
    $N_{\rm scalar} \gg 1$.
    If one generalizes the scalar sector of the standard model to include
    a large number $N$ of complex scalar fields, with a global U$(N)$ symmetry,
    then one may expand phsyical results in powers of $1/N$ and compare
    the resulting large-$N$ predictions with those of the $\epsilon$-expansion.
    In brief, the result is that the $\epsilon$-expansion does {\em not\/}
    work well when $N \gg 1$.
    However, the $\eps$-expansion alerts one to its own failure
    by producing next-to-leading order corrections that are
    significantly larger than the leading-order result when $\epsilon=1$.
    For example, the ``tricritical slope''%
    \footnote
	{%
	When $N$ is sufficiently large, an infrared stable fixed point
	and a tricritical fixed point appear.
	If $\lambda / g^2$ is sufficiently large then the theory flows to the
	stable fixed point (and thus has a second order transition).
	If $\lambda / g^2$ is sufficiently small then the theory undergoes
	a first order phase transition.
	The tricritical slope is the slope of the line separating the
	two domains.
	}
    has the asymptotic forms
    \begin {eqnarray}
       {\lambda\over q^2}
    &=&
       {3 \over N} \left[ \, 54 -126 \, \eps + O(\eps^2)\right]
       + O\left({1 \over N^2}\right) \,.
    \label {tricrit-eps}
    \\
    &=&
       {3 \over N} \left ({96 \over \pi^2} \right)
       + O\left({1 \over N^2}\right) .
    \label {tricritical}
    \end {eqnarray}
    The lowest order $\epsilon$-expansion result, $162/N$, differs
    from the correct large-$N$ result (\ref {tricritical})
    by more than a factor of 5, but the expansion (\ref {tricrit-eps})
    is obviously poorly behaved and unreliable at $\epsilon = 1$.
    \end {itemize}

Altogether, the available information suggests that the $\epsilon$-expansion
can be a useful approximation for the standard model
(or other gauge theories containing a small number of scalar fields).
Most importantly, the $\epsilon$-expansion predicts that
the bounds for viable electroweak baryogenesis are even more stringent than
suggested by a one-loop analysis in three dimensions.
Clearly, calculations of additional physical quantities at next-to-leading
order in the $\epsilon$-expansion should be performed to further
confirm the reliability of the method.

\end {document}